\documentclass[pra,showpacs,twocolumn,eqsecnum,superscriptaddress]{revtex4}
\usepackage{makeidx}
\usepackage{eurosym}
\usepackage{amssymb}
\usepackage{amsfonts}
\usepackage{amsmath}
\usepackage{setspace}
\usepackage[english]{babel}
\usepackage{graphicx}
\usepackage[applemac]{inputenc}
\usepackage{color}
\usepackage{ifthen}
\usepackage{float}
\usepackage{verbatim}
\usepackage{subfigure}

\begin{document}
\title{Quantum phase transitions for two coupled sites with
dipole-coupled effective Jaynes-Cummings model}
\author{Lei Tan}
\affiliation{Institute of Theoretical Physics, Lanzhou University,
Lanzhou 730000, China} \affiliation{Beijing National Laboratory
for Condensed Matter Physics, Institute of Physics, Chinese
Academy of Sciences, Beijing 100190, China}
\author{Yu Qing Zhang}
\affiliation{Institute of Theoretical Physics, Lanzhou University,
Lanzhou 730000, China}
\author{Wu Ming Liu}
\affiliation{Beijing National Laboratory for Condensed Matter
Physics, Institute of Physics, Chinese Academy of Sciences,
Beijing 100190, China}
\date{\today}

\begin{abstract}
The nature of the ground  states for a system composed of two
coupled cavities with each containing a pair of dipole-coupled
two-level atoms are studied over a wide range of detunings and
dipole coupling strengths. The cases for three limits of  exact
resonance, large positive and negative  detunings  are discussed,
and four types  of the ground states are revealed.  Then  the
phase diagrams of the ground state are plotted by choosing three
different ``order parameters". We find that  the phase space,
determined by the combinative action of detuning  and the dipole
coupling strength, is divided  into four regions. This is
different from the general Bose-Hubbard model and more richer
physics are presented in the two-site coupled cavities system.
That is, the insulator region may be polaritonic or atomic and the
superfluid region may be polaritonic or photonic in nature.

\end{abstract}

\maketitle

\section{Introduction}

The simulation of the strongly correlated many-body systems
described by Bose-Hubbard model has received great advances in
optical
lattices\cite{Greiner,Bloch,Jaksch,Albus,Scarola,Christoph,Spielman,Schneider,Strohmaier,Compton}
and coupled-cavity
systems\cite{Hartmann1,Greentree,Angelakis,Hartmann2,Rossini,Aichhorn,Cho,Carusotto,Schmidt0,Jens,Hartmann3,Tomadin,Tomadin2,Ciccarello}.
Both of them depend on the competition between the local
interaction and the nonlocal tunneling, but there are also some
differences for this two basic models. In optical lattices, the
quantum phase transitions (QPT) in a gas of ultracold atoms with
periodic potentials is described by the Bose-Hubbard model with on
site two atoms interacting and hopping between the adjacent sites.
However, in the coupled-cavity systems, two types of particles are
involved to study the many-body dynamics and  its realization
relays on the strong light-matter coupling regime. Therefore, the
QPT is due to the transferring of the excitations from polaritonic
to photonic rather than purely bosonic or purely fermionic
entities. However, the  realization of the strong coupling  in
experiment is the greatest bottleneck for the QPT manipulated in
coupled-cavity system. To be optimistic, with the progress in the
realization of strong light-matter coupling regime in both
atomic\cite{Raimond,Birnbaum} and
solid-state\cite{Reithmaier,Hennessy} cavity quantum
electrodynamics devices with single two-level emitters in high-Q
resonators, the QPT  for coupled-cavity arrays of
Jaynes-Cummings(JC) model systems and its variants attracts more
and more attentions.

Subsequent works that deal with coupled nonlinear cavities arrays
have addressed the dynamics in the two coupled cavities for its
great freedom and flexibility, which can provides a convenience
controllable platform for engineering the transport of quantum
states via photonic processes and the exact numerical solutions
can be easily found and then some analytical approximations can be
used. Only very recently the investigation of such systems of two
coupled cavities has begun. The quantum states
transfers\cite{Irish}, the atomic state transfer\cite{Ogden}, the
quantum phase gates\cite{zheng}, the bipartite entanglement
entropies\cite{Irish1}, the one-excitation dynamics\cite{zhang},
the photon correlations\cite{Ferretti}, the time evolution of the
population imbalance\cite{Schmidt}, the photonic tunneling
effect\cite{Guo} and the emission characteristics\cite{Kanp} have
been studied. To the best of our knowledge, most of the previous
works focused on this system are limited to the single atom-cavity
interactions without consideration of an additional interatomic
coupling. With the advances of  the technology, the interatomic
dipole-dipole interaction may be realized in several solid-state
systems, such as an ensemble of quantum dots\cite{Scheibner} and
Bose-Einstein condensates~\cite{Schneble}. At present, the
stationary entanglement\cite{Nicolosi,Wang} and QPT\cite{Li}  for
dipole-coupled two-level atoms in single-mode cavity have been
investigated  theoretically. Then an investigation of the QPT to
the system of two coupled cavities with dipole-dipole interaction
is also of considerable significance and highly called for.

The previous work\cite{Irish} has identified some of the unique
features of  QPT in the  coupled two-site JC model. The nature of
the ground states for the system can be divided into four types
corresponding to different parameter values of atom-field
detunings and cavity-cavity hopping strengths. Differing from the
Bose-Hubbard model, the insulator state may be either atomic or
polaritonic, while the superfluid state may be photonic or
polaritonic in nature. In this paper, we extend the
work\cite{Irish} to each cavity containing two dipole-coupled
atoms. We find that in the presence of weak cavity-cavity
coupling, the effective on-site repulsion not only attributes to
the  atom-photon interaction and the detuning, but also depends on
the interatomic dipole-coupled strength. So the atomic
dipole-dipole  interaction provides an additional parameter, and
more importantly, richer physics for the system.

The paper is organized as follows. We first describe the model
under consideration and then simplify it to an effective form,
with reformed atomic energy and atom-field coupling strength. We
then address the nature of the ground state of the system under a
wide range of detuning and dipole coupling strengths. Thirdly, we
discuss the quantum phase transitions of the system by choosing
three different ``order parameters". We finally present our
conclusions.

\section{MODEL}

The system under consideration  consists of two identical
single-mode cavities with each cavity containing  two coupled
two-level atoms  through  dipole-dipole interaction. The two
cavities are coupled  by  hopping strength $A$, therefore the
photons may hop between them. Besides, the model is ideal without
taking into account the dissipation induced by  atomic spontaneous
emission and photonic escape from the cavities. In such a case,
the Hamiltonian for the  coupled two-cavity system is given by
$(\hbar=1)$
\begin{eqnarray}
H&=&H_{1}+H_{2}+H_{12},\nonumber\\
H_{1}&=&\omega_{c}a^{\dagger}_{1}a_{1}+\sum_{i=1,2}[\omega_{a}\sigma_{i}^{\dag}\sigma_{i}+g(a^{\dagger}_{1}\sigma_{i}+a_{1}\sigma_{i}^{\dag})]\nonumber\\
&+&J(\sigma_{1}^{\dag}\sigma_{2}+\sigma_{1}\sigma_{2}^{\dag}),\nonumber\\
H_{2}&=&\omega_{c}a^{\dagger}_{2}a_{2}+\sum_{j=3,4}[\omega_{a}\sigma_{j}^{\dag}\sigma_{j}+g(a^{\dagger}_{2}\sigma_{j}+a_{2}\sigma_{j}^{\dag})]\nonumber\\
&+&J(\sigma_{3}^{\dag}\sigma_{4}+\sigma_{3}\sigma_{4}^{\dag}),\nonumber\\
H_{12}&=&A(a^{\dagger}_{1}a_{2}+a_{1}a^{\dagger}_{2}),
\end{eqnarray}
where $\omega_{a}$ and $\omega_{c}$ are the resonance frequencies
for atoms and cavities, respectively. Supposing the same
atom-field coupling strength in the system, it is   defined by an
uniform parameter $g$. $a^{\dagger}_{i}$ and $a_{i}$ are the
creation and annihilation operators of the  field in cavity $i$
($i=1,2$). $\sigma^{+}_{j}$ and  $\sigma_{j}$ represent the atomic
raising and lowering  operators of the atom $j$ ($j=1,2,3,4$). For
static atoms,  the coherent dipole-dipole  interaction between
them can be  given by
\begin{eqnarray}
J=|\textbf{d}|^{2}(1-3\cos^{2}\theta)/\textbf{r}_{12}^{3},
\end{eqnarray}
where, $\textbf{r}_{12}$=$\textbf{r}_{1}-\textbf{r}_{2}$ is the
distance  between the two  atoms located at $\textbf{r}_{1}$ and
$\textbf{r}_{2}$. $\theta$ is  the angle between $\textbf{r}_{12}$
and  the atomic dipole moment $\textbf{d}$. Here, we assume  the
dipole moments of the two atoms  are parallel to each other and
are polarized in the direction perpendicular to the interatomic
axis. Then, $J$ can be simplified as
\begin{eqnarray}
J=|\textbf{d}|^{2}/\textbf{r}_{12}^{3},
\end{eqnarray}
and  its strength  can be adjusted by changing the positions of
the two atoms in each cavity. $H_{1}$ and $H_{2}$ show the
atom-field interaction and the atom-atom  coupling  in each site,
respectively. The cavity-cavity coupling is depicted by $H_{12}$.
The total excitation  for the Hamiltonian $H$ can be defined as
$N=a^{\dagger}_{1}a_{1}+a^{\dagger}_{2}a_{2}+\sigma_{1}^{\dag}\sigma_{1}+\sigma_{2}^{\dag}\sigma_{2}$.
We assume the total number of excitations $N$ is conserved and
exactly two excitations in the system. Then  Hamiltonian $H_{1}$
and $H_{2}$ can be transformed into two simple forms by unitary
transformation\cite{Nicolosi},
\begin{eqnarray}
H_{1eff}&=&U^{\dagger}_{1}H_{1}U_{1}=\omega_{c}a^{\dagger}_{1}a_{1}+(\omega_{a}+J)\sigma_{1}^{\dag}\sigma_{1}\nonumber\\
&+&\sqrt{2}g(a^{\dagger}_{1}\sigma_{1}+a_{1}\sigma_{1}^{\dag})],\nonumber\\
H_{2eff}&=&U^{\dagger}_{2}H_{1}U_{2}=\omega_{c}a^{\dagger}_{2}a_{2}+(\omega_{a}+J)\sigma_{3}^{\dag}\sigma_{3}\nonumber\\
&+&\sqrt{2}g(a^{\dagger}_{2}\sigma_{3}+a_{2}\sigma_{3}^{\dag})].
\end{eqnarray}
where,
$U_{1}=\exp[-\frac{\pi}{4}(\sigma^{\dagger}_{1}\sigma_{2}+\sigma^{\dagger}_{2}\sigma_{1})]$,
$U_{2}=\exp[-\frac{\pi}{4}(\sigma^{\dagger}_{3}\sigma_{4}+\sigma^{\dagger}_{4}\sigma_{3})]$.
In the transformed form, the dipole coupled atoms are denoted by
two fictitious  atoms. Only one of them couples to the field mode
with  frequencies $\omega_{a}+J$, but the other atom freely
evolves  decoupling from the field. The effective  coupling
strength also changes from $g$ to $\sqrt{2}g$. In this paper, we
pay our attention to the  strong atom-cavity coupling regime which
can be put into practice only when $A\ll g$. On this condition,
the eigenstates of the individual cavity should be expressed  by
the  dressed states
\begin{eqnarray}
|0_{i}\rangle&=&|g_{i}\rangle|0_{i}\rangle,\nonumber\\
|n^{-}_{i}\rangle&=&\sin\frac{\theta_{n}}{2}|e_{i}\rangle|(n-1)_{i}\rangle-\cos\frac{\theta_{n}}{2}|g_{i}\rangle|n_{i}\rangle,\nonumber\\
|n^{+}_{i}\rangle&=&\cos\frac{\theta_{n}}{2}|e_{i}\rangle|(n-1)_{i}\rangle+\sin\frac{\theta_{n}}{2}|g_{i}\rangle|n_{i}\rangle.
\end{eqnarray}
where $i=1,2$ indicates the cavity number, $\sqrt{n}$ is a photon
number state, $\theta_{n}=\arctan 2\sqrt{2}g\sqrt{n}/(\Delta+J)$,
and $\Delta=\omega_{a}-\omega_{c}$ is the detuning between the
atom and the field. The  eigenenergies of these eigenstates are
\begin{eqnarray}
E_{i}^{0}&=&0,\nonumber\\
E_{i}^{n\mp}&=&n\omega_{c}+\frac{\Delta+J}{2}\mp\frac{1}{2}\sqrt{(\Delta+J)^{2}+8g^{2}n}.
\end{eqnarray}

The effective form of Hamiltonian $H$ can be rewritten as
\begin{eqnarray}
H_{eff}=H_{1eff}+H_{2eff}+H_{12}
\end{eqnarray}

Because there are only two  excitations in the system, we can
write out the the corresponding state of $H_{eff}$ in the  order
of increasing energy and divide them into five groups, defined as
$|\phi_{1}\rangle$, $|\phi_{2}\rangle$, $|\phi_{3}\rangle$,
$|\phi_{4}\rangle$, $|\phi_{5}\rangle$, corresponding to subspaces
$\{|1^{-}_{1}\rangle\otimes|1^{-}_{2}\rangle\}$,
$\{|2^{-}_{1}\rangle\otimes|0_{2}\rangle,
|0_{1}\rangle\otimes|2^{-}_{2}\rangle\}$,
$\{|1^{-}_{1}\rangle\otimes|1^{+}_{2}\rangle,
|1^{+}_{1}\rangle\otimes|1^{-}_{2}\rangle\}$,
$\{|2^{+}_{1}\rangle\otimes|0_{2}\rangle,
|0_{1}\rangle\otimes|2^{+}_{2}\rangle\}$,
$\{|1^{+}_{1}\rangle\otimes|1^{+}_{2}\rangle\}$, respectively.
Obviously, the energy difference between the adjacent subspaces
depends on the parameter values in  $H_{eff}$. Then the
probability distribution of the ground states in the five
subspaces, as well as its nature,  will be different for taking
different parameter values.

\begin{figure}[ptb]
\centerline{\includegraphics*[width=0.8\textwidth]{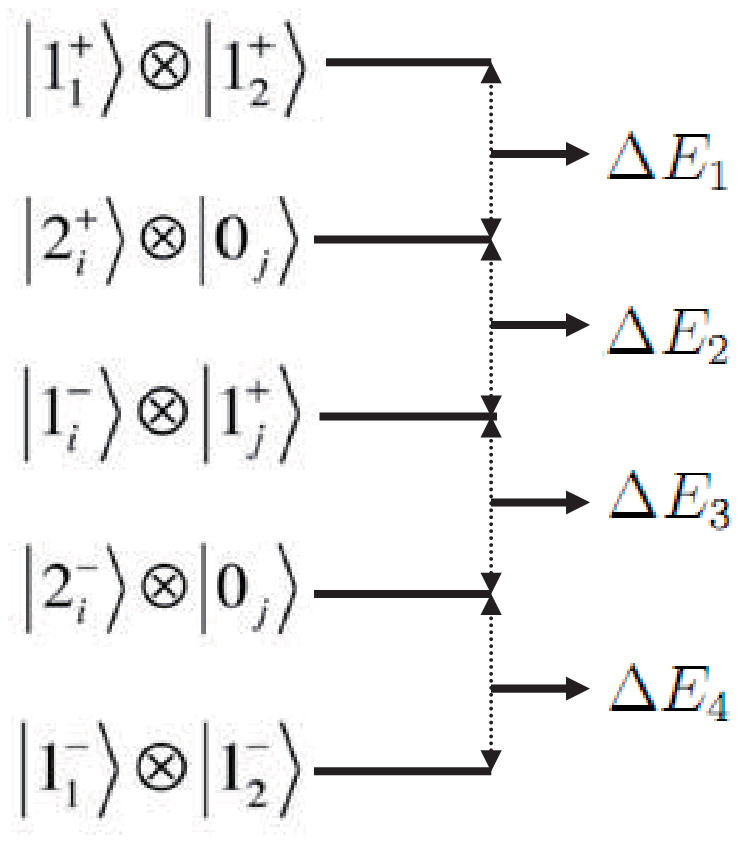}}\caption{
Energy levels for the two-cavity system in the absence of hopping
(A=0), $\Delta E_{i}$ $(i=1,2,3,4)$ is the energy gap
between the adjacent energy levels.}%
\label{Fig1}%
\end{figure}

\begin{figure}[ptb]
\centerline{\includegraphics*[width=0.5\textwidth]{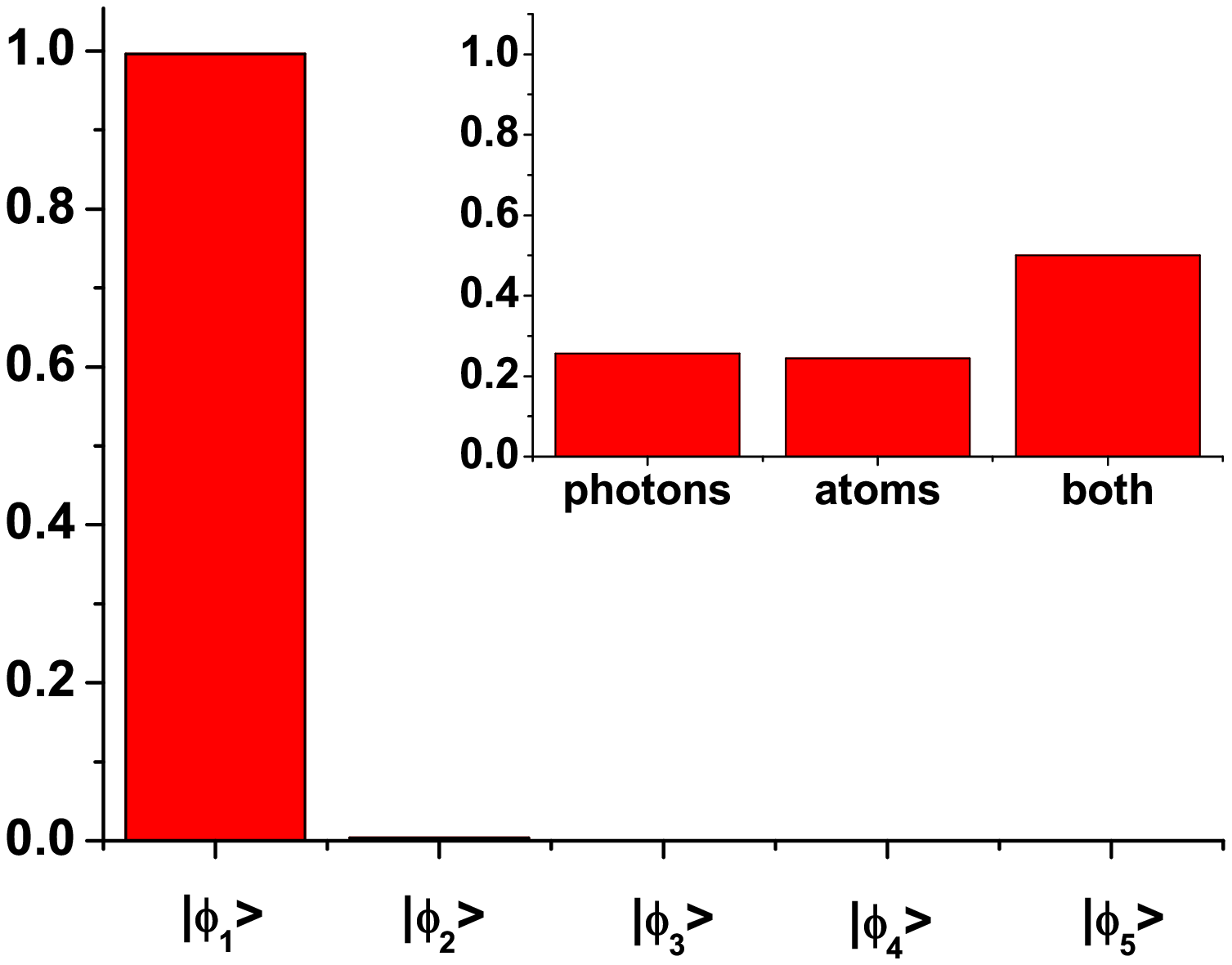}}\caption{(Color
online) Occupation probabilities of the ground state in the five
subspaces. The inset shows the probability distribution of
excitations among states with purely photonic, purely atomic, and
mixed of them. The hopping strength between the two cavities is
weak for  $A=0.1g$, the detuning between the atom and the field
is $\Delta=0$, and the atom-atom coupling strength is $J=0.1g$. }%
\label{Fig2}%
\end{figure}

\begin{figure}[ptb]
\centerline{\includegraphics*[width=0.5\textwidth]{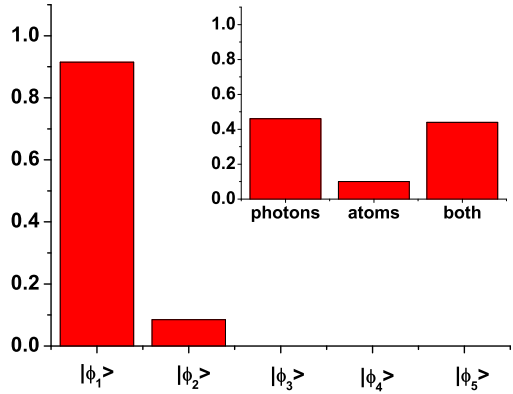}}\caption{(Color
online) Occupation probabilities of the ground state in the five
subspaces. The inset shows the probability distribution of
excitations among states with purely photonic, purely atomic, and
mixed of them. The hopping strength between the two cavities is
weak for  $A=0.1g$, the detuning between atom and field is
$\Delta=0$, and the atom-atom coupling strength is $J=g$. }%
\label{Fig2}%
\end{figure}

\section{The nature of the ground state}
In the weak  cavity-cavity coupling limit, we can understand the
nature of the ground state by considering the effects of
parameters in the effective Hamiltonian in Eq.(4). In fact, the
energy gap $\Delta E_{i}$ $(i=1,2,3,4)$ between the five subspaces
not only dependents on the atom-field coupling strength $g$ and
detuning $\Delta$, but also relies  on the atom-atom coupling
strength $J$, which is shown in Fig.1.

where
\begin{eqnarray}
\Delta E_{1}&=&\sqrt{(\Delta+J)^{2}+8g^{2}}\nonumber\\
&-&\frac{1}{2}[\sqrt{(\Delta+J)^{2}+16g^{2}}+(\Delta+J)],\nonumber\\
\Delta E_{2}&=&\frac{1}{2}[(\Delta+J)+\sqrt{(\Delta+J)^{2}+16g^{2}}],\nonumber\\
\Delta E_{3}&=&\frac{1}{2}[\sqrt{(\Delta+J)^{2}+16g^{2}}-(\Delta+J)],\nonumber\\
\Delta E_{4}&=&\sqrt{(\Delta+J)^{2}+8g^{2}}\nonumber\\
&-&\frac{1}{2}[\sqrt{(\Delta+J)^{2}+16g^{2}}-(\Delta+J)].
\end{eqnarray}

To illustrate the behavior of the ground state operating in the
dipole-dipole interaction, we first pay our attention to the
resonant condition $\Delta=0$, and  three cases for $J\ll g$,
$J\approx g$, and $J\gg g$ will be discussed in the following.

At $J\ll g$, there is a large energy gap between
$|\phi_{1}\rangle$ and  $|\phi_{2}\rangle$ due to  the photon
blockade effect, for which the presence of one photon in the
cavity blocks the entering of the subsequent photons. On this
condition, the ground state of the system is approximately
$|1^{-}_{1}\rangle\otimes|1^{-}_{2}\rangle$, as shown in Fig.2.
Moreover, $|1^{-}\rangle$ is nearly the maximal entanglement state
of atom and field. For this state, only one excitation in each
cavity  with almost equal probabilities for   atomic and field
excitations. This can also be confirmed from the inset of Fig.2.
Thus the ground state is a polaritonic insulatorlike state, which
is analogous to the Mott insulator state in the Bose-Hubbard
model.
\begin{figure}[ptb]
\centerline{\includegraphics*[width=0.5\textwidth]{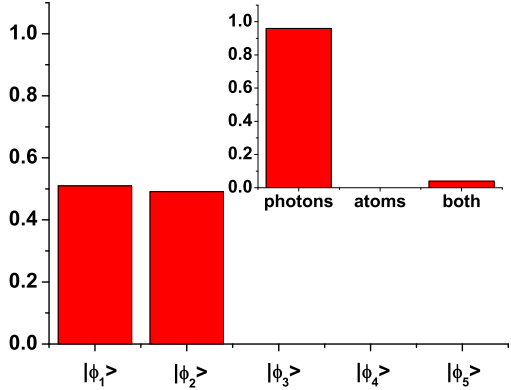}}\caption{(Color
online) (Occupation probabilities of the ground state in the five
subspaces. The inset shows the probability distribution of
excitations among states with purely photonic, purely atomic, and
mixed of them. The hopping strength between the two cavities is
weak for  $A=0.1g$, the detuning between atom and field is
$\Delta=0$, and the atom-atom coupling strength is $J=10g$. }%
\label{Fig2}%
\end{figure}

\begin{figure}[ptb]
\centerline{\includegraphics*[width=0.5\textwidth]{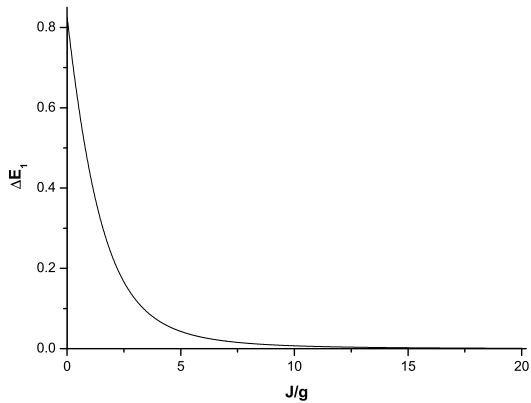}}\caption{
The energy gap $\Delta E_{1}$ versus dipole coupling strength $J$
at $\Delta=0$. The hopping
strength between the two cavities is weak for  $A=0.1g$. }%
\label{Fig3}%
\end{figure}

\begin{figure}[ptb]
\centerline{\includegraphics*[width=0.5\textwidth]{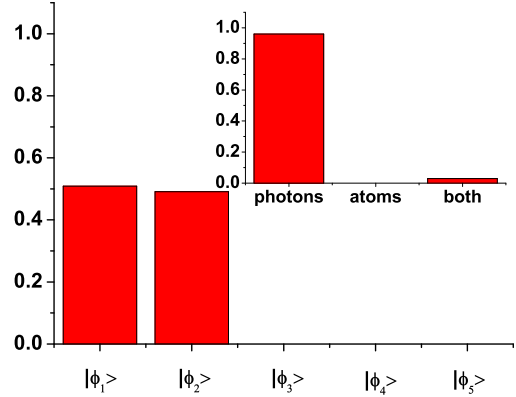}}\caption{(Color
online) Occupation probabilities of the ground state in the five
subspaces. The inset shows the probability distribution of
excitations among states with purely photonic, purely atomic, and
mixed of them. For large positive atom-field detuning, the effects
of the dipole coupling strength can be ignored. Here, we  choose
$A=0.1g$,  $\Delta=10g$ and
$J=g$. }%
\label{Fig4}%
\end{figure}

\begin{figure}[ptb]
\centerline{\includegraphics*[width=0.5\textwidth]{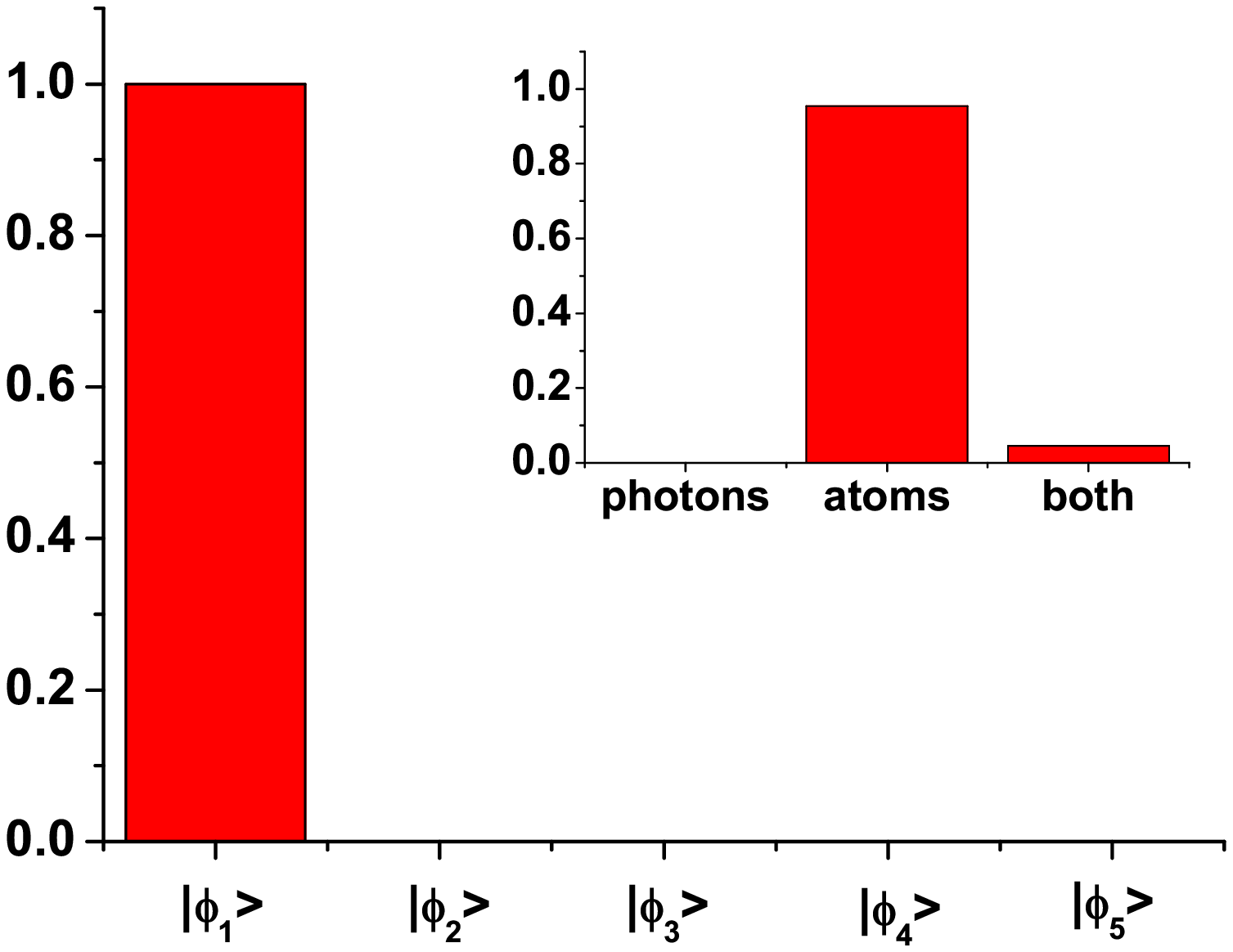}}\caption{(Color
online) Occupation probabilities of the ground state in the five
subspaces. The inset shows the probability distribution of
excitations among states with purely photonic, purely atomic, and
mixed of them. For large negative atom-field detuning, two cases
of weak and strong dipole coupling strength are shown. Here, we
choose $A=0.1g$  and $\Delta=-10g$. Other parameter value is
$J=0.1g$. }%
\label{Fig4}%
\end{figure}

\begin{figure}[ptb]
\centerline{\includegraphics*[width=0.5\textwidth]{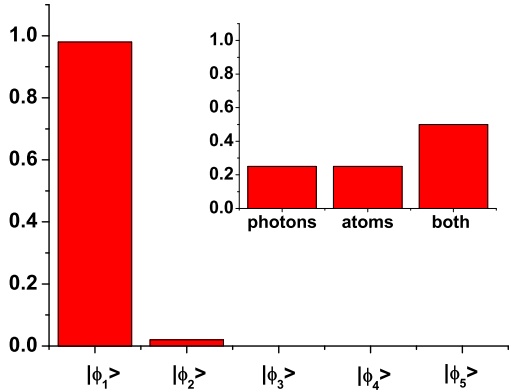}}\caption{(Color
online) Occupation probabilities of the ground state in the five
subspaces. The inset shows the probability distribution of
excitations among states with purely photonic, purely atomic, and
mixed of them. For large negative atom-field detuning, two cases
of weak and strong dipole coupling strength are shown. Here, we
choose $A=0.1g$  and $\Delta=-10g$. Other parameter value is
$J=10g$. }%
\label{Fig4}%
\end{figure}
Compared to Fig.2,  Figure 3  indicates  a more different behavior
for $J\approx g$. Besides $|\phi_{1}\rangle$, the subspace
$|\phi_{2}\rangle$ is also occupied for the ground state of the
system. From its inset, we find that both  the atom and the field
are excited,  indicating a polaritonic superfluidlike state.

When  $J\gg g$, the ground state occupies the subspaces
$|\phi_{1}\rangle$ and $|\phi_{2}\rangle$ with  almost same
probabilities, as represented in Fig.4. However, only the photons
are excited in this state. This is because, in this limit,
$|n^{-}\rangle\approx-|gn\rangle$, so the ground state is a
delocalized photon state in nature. The state of this form is a
photonic superfluidlike state.

The results can be understood easily. Similar to the effect of
detuning in Ref.\cite{Irish}, at $\Delta=0$, the  energy gap of
$\Delta E_{1}$ is a monotonic decreasing function of $J$, see
Fig.5. With the increase of $J$, the photon blockade effect is
destroyed, leading to an increase of the occupied probability of
$|\phi_{2}\rangle$. When $J\approx 10g$, $\Delta E_{1}$ is almost
zero, so the subspaces $|\phi_{1}\rangle$ and $|\phi_{2}\rangle$
nearly  degenerate. Oppositely, $\Delta E_{2}$ is a monotonic
increasing function of $J$ with a large initial value. Thus, the
subspace $|\phi_{3}\rangle$ can not be occupied.

Next, we discuss the case of large positive detuning. At
$\Delta=0$, the nature of the ground state changes with different
values of $J$. However, when $\Delta\gg g$, no matter  what value
of $J$, the energy gap of $\Delta E_{1}$ is always zero. While
$\Delta E_{2}$ is a monotonic increasing function of $\Delta+J$.
Even $J=0$, the energy gap  $\Delta E_{2}$ is very large.
Therefore, the ground state has equal occupation  probabilities in
the subspaces $|\phi_{1}\rangle$ and $|\phi_{2}\rangle$, but zero
probability in other subspaces. More interestingly, in the large
positive detuning limit, the superposition coefficients of the
dressed states in Eq.(5) have particular values,
$\sin\frac{\theta_{n}}{2}$$\approx0$ and
$\cos\frac{\theta_{n}}{2}$$\approx1$. Then,
$|n^{-}\rangle\approx-|g\rangle|n\rangle$. Consequently, not only
the constitution of the ground state but also its nature is fixed.
In Fig.6, the inset shows that the excitation is  photonic rather
than  atomic. This is  because the energy of the atoms is larger
than that of the photons when $\Delta>0$. Analogous to the
situation shown in Fig.4, the ground state indicates  photonic
superfluidlike nature.

At $\Delta<0$, the energy of  atoms is  smaller than that of
photons. In the limit of $-\Delta\gg g,J$, we find $\Delta
E_{1}\approx|\Delta|$, the ground state is approximate
$|1^{-}_{1}\rangle\otimes|1^{-}_{2}\rangle$. In this condition,
$\sin\frac{\theta_{n}}{2}$$\approx1$ and
$\cos\frac{\theta_{n}}{2}$$\approx0$. Thus,
$|1^{-}\rangle\approx|e\rangle|0\rangle$, The excitation is almost
atomic rather than photonic, standing  for an atomic insulatorlike
state. The result is illustrated in Fig.7. However, when $J$
approaches to the value of $-\Delta$, the value of $|\Delta+J|$
decreases. That is, the energy difference between atom and photon
is less and  less. As a result, the photonic and atomic
excitations coexist, as shown in  Fig.8, corresponding to the
polaritonic insulatorlike state.

\section{Phase diagrams}

 The phase diagrams of these states can be distinguished using
the  corresponding  ``order parameters". In superfluid states, the
excitations in each cavity is uncertain, resulting in a nonzero
variance of the total excitation number $\Delta N_{1}$.
Oppositely, in the insulator state the number of excitations per
cavity is constant and thus has zero variance. However, the
insulator state may be either atomic or polaritonic and the
superfluid state may be photonic or polaritonic in nature. To
determine the  allowed  types of particles involved in the state,
the atomic excitation number variance $\Delta N_{A1}$ should be
taken as the ``order parameter". $\Delta N_{A1}=0$ is
corresponding to the atomic insulator state or the photonic
superfluid state, while $\Delta N_{A1}>0$ revealing  the
polaritonic nature. Only when  $\Delta N_{1}\Delta N_{A1}>0$, it
shows the polaritonic superfluid characteristic for the ground
state of the system.

\begin{figure}[ptb]
\centerline{\includegraphics*[width=0.5\textwidth]{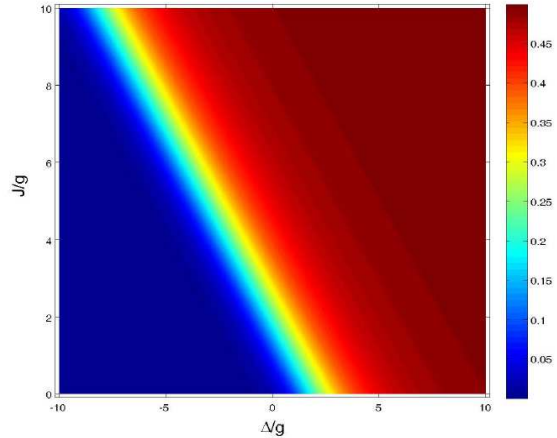}}\caption{(Color
online) For  weak hopping strength $A=0.1g$, $\Delta N_{1}$ is
plotted versus atom-field detuning $\Delta$ and dipole coupling
strength $J$
 in the ground state of the coupled two-site and  two-excitation  system. }%
\label{Fig4}%
\end{figure}

\begin{figure}[ptb]
\centerline{\includegraphics*[width=0.5\textwidth]{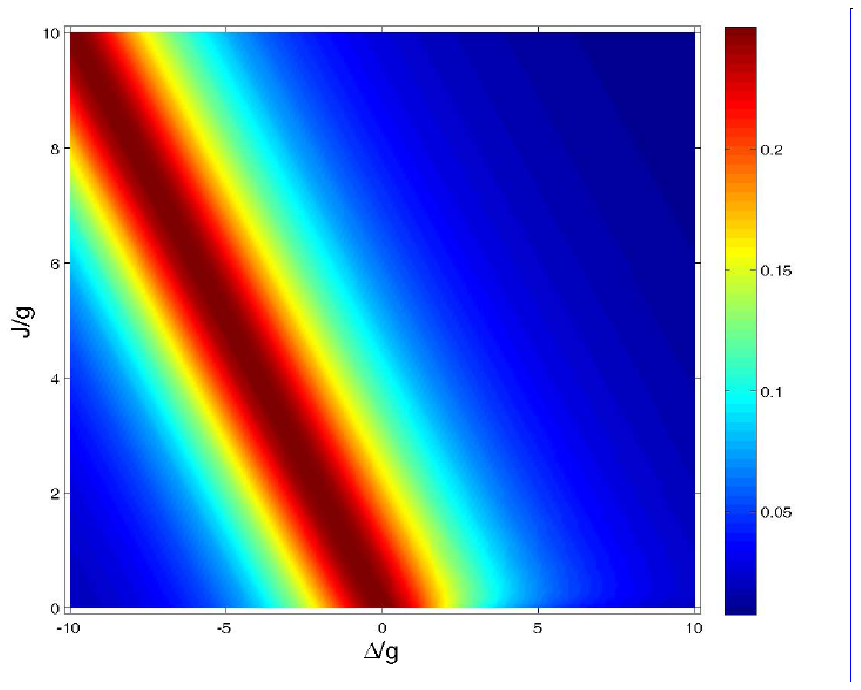}}\caption{(Color
online) For  weak hopping strength $A=0.1g$, $\Delta N_{A1}$ is
plotted versus atom-field detuning $\Delta$ and dipole coupling
strength $J$
 in the ground state of the coupled two-site and  two-excitation  system. }%
\label{Fig4}%
\end{figure}

\begin{figure}[ptb]
\centerline{\includegraphics*[width=0.5\textwidth]{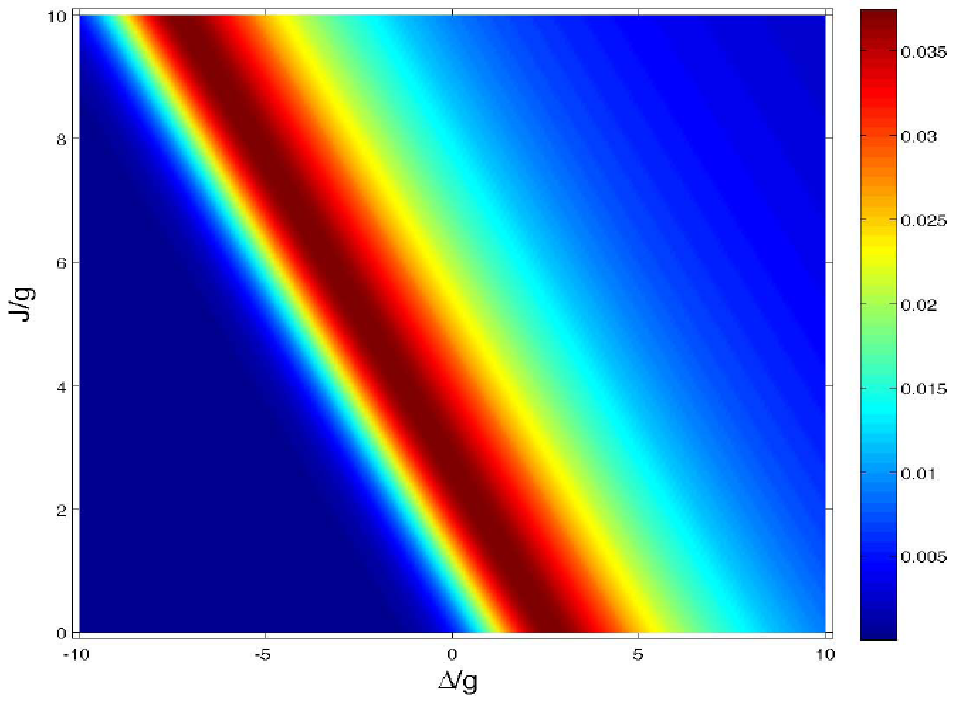}}\caption{(Color
online) For weak hopping strength $A=0.1g$, $\Delta
N_{1}\cdot\Delta N_{A1}$ is plotted versus atom-field detuning
$\Delta$ and dipole coupling strength $J$
 in the ground state of the coupled two-site and  two-excitation  system. }%
\label{Fig4}%
\end{figure}

To begin with, we  should distinguish the insulator and superfluid
areas in the phase diagram, which are determined by the  total
excitation number $\Delta N_{1}$. With  Eq.(4), we can give out
the expression of  $\Delta N_{1}$  directly.
\begin{eqnarray}
N_{1}&=&a^{\dagger}_{1}a_{1}+\sigma_{1}^{\dag}\sigma_{1},\nonumber\\
\Delta N_{1}&=&\langle N^{2}_{1}\rangle-\langle N_{1}\rangle^{2}.
\end{eqnarray}
In Fig.9, $\Delta N_{1}$ is plotted  under a wide range values of
the atom-cavity detunings $\Delta$ and the dipole-dipole
interaction strength  $J$.
 It is apparent that the phase diagram  is divided into two sections.
The insulator region is under the boundary  where $\Delta<0$ while
above the boundary is the  superfluid region. There is also an
area which is  symmetric to the insulator area, where the $\Delta
N_{1}$ with a maximum value 0.5, indicating a most evident
superfluidity.

To find out the polaritonic area, we should introduce the second
``order parameters" $\Delta N_{A1}$, where
\begin{eqnarray}
N_{A1}&=&a^{\dagger}_{1}a_{1}+\sigma_{1}^{\dag}\sigma_{1},\nonumber\\
\Delta N_{A1}&=&\langle N^{2}_{A1}\rangle-\langle
N_{A1}\rangle^{2}.
\end{eqnarray}
In Fig.10, we find that the polaritonic area   approximately
spreads at both sides of the   line  $J=-\Delta$. The more closer
to the  line the more obvious  the   polaritonic nature.  In fact,
when $J=-\Delta$, $\Delta E_{1}=(2\sqrt{2}-2)g$, it is  the
conditions for  photon blockade effect obviously. Then the ground
state only occupies the subspace $|\phi_{1}\rangle$, with
$\sin\frac{\theta_{n}}{2}=\cos\frac{\theta_{n}}{2}=\frac{1}{2}$,
standing  for a maximal entanglement state of the atom and the
field. So the ground state indicates  polaritonic insulatorlike
nature.

So far we may  guess that on the bases of what are embodied in
Fig.9 and Fig.10, there must be some overlapped areas in the two
figures. In Fig.11,  the product of  $\Delta N_{1}$ and $\Delta
N_{A1}$ is shown as a contour plot. We can clearly identify that
there is an area not only in the superfluid region in Fig.9, but
also in the polaritonic area in Fig.10, representing polaritonic
superfluidlike characters.

Then it is obvious that the phase space in Fig.9 is divided into
four sections,  from left bottom to top right corner, in order of
atomic insulatorlike region, polaritonic insulatorlike region,
polaritonic superfluidlike region and photonic superfluidlike
region.

\section{Conclusions}

In summary, we have investigated the QPT of a system composed of
two coupled cavities, each containing a pair of two-level atoms
with dipole-dipole interaction. In the conditions of fixed
cavity-cavity interaction and atom-cavity coupling strength, the
nature of the ground state is dependent on the constituents of the
dressed states  in each cavity and the occupation probabilities of
the ground state  in the five subspaces. Moreover, both of them
attribute to the dipole-dipole interaction strength  between the
localized atoms and  the atom-field detuning in each cavity. By
choosing three different order parameters, we found that the
ground state of the system  represented more richer behaviors than
the Bose-Hubbard model. Four  types of states are revealed, which
divide the phase space into four regions. They are the atomic
insulatorlike state, the polaritonic insulatorlike state, the
polaritonic superfluidlike state and the photonic superfluidlike
state. In the scope of parameter values we taken in this paper,
the  insulator or superfluid phases is determined by the
combinative effect of $\Delta$ and $J$, that is the value of
$\Delta+J$. Small negative values of it is in favour of
polaritonic insulatorlike  states while for small positive value
of it embodies polaritonic superfluidlike state. The more larger
negative value of $\Delta+J$, the more obvious of atomic
insulatorlike  nature, and oppositely it shows photonic
superfluidlike nature.

\begin{acknowledgments}
This work was supported by NSFC under grants Nos. 10704031,
10874235, 10934010 and 60978019, the NKBRSFC under grants Nos.
2009CB930701, 2010CB922904 and 2011CB921500, and FRFCU under grant
No. lzujbky-2010-75.
\end{acknowledgments}


\begin{thebibliography}{99}
\bibitem{Greiner} M. Greiner, et al., Nature (London) \textbf{415}, 39 (2002).
\bibitem{Bloch} I. Bloch, J. Dalibard, and W. Zwerger, Rev. Mod. Phys. 80, 885 (2008).
\bibitem{Jaksch} D. Jaksch, C. Bruder, J. I. Cirac, C. W. Gardiner, and P. Zoller, Phys. Rev. Lett. \textbf{81}, 3108 (1998).
\bibitem{Albus}A. Albus, F. Illuminati, and J. Eisert, Phys. Rev. A \textbf{68}, 023606 (2003).
\bibitem{Scarola} V. W. Scarola and S. Das Sarma, Phys. Rev. Lett. \textbf{95}, 033003 (2005).
\bibitem{Christoph} C. Maschler and H. Ritsch, Phys. Rev. Lett. \textbf{95}, 260401 (2005).
\bibitem{Spielman}  I. B. Spielman, W. D. Phillips, and J. V. Porto, Phys. Rev. Lett. \textbf{98}, 080404 (2007).
\bibitem{Schneider}U. Schneider, et al., Science \textbf{322}, 1520 (2008).
\bibitem{Strohmaier}R. J$\ddot{o}$rdens, et al., Nature (London) \textbf{455}, 204 (2008).
\bibitem{Compton} K. Jim\'{e}nez$-$Garc\'{\i}a, et al., Phys. Rev. Lett. \textbf{105}, 110401 (2010).


\bibitem{Hartmann1}M. J. Hartmann, F. G. S. L. Brand$\tilde{a}$o, and M. B. Plenio, Nat. Phys. \textbf{2}, 849 (2006).
\bibitem{Greentree}A. D. Greentree, C. Tahan, J. H. Cole, and L. C. L. Hollenberg, Nat. Phys. \textbf{2}, 856 (2006).
\bibitem{Angelakis}D. G. Angelakis, M. F. Santos, and S. Bose, Phys. Rev. A \textbf{76}, 031805(R) (2007).
\bibitem{Hartmann2}M. J. Hartmann, F. G. S. L. Brand$\tilde{a}$o, and M. B. Plenio, Laser Photon. Rev. \textbf{2}, 527 (2008).
\bibitem{Rossini}D. Rossini and R. Fazio, Phys. Rev. Lett. \textbf{99}, 186401 (2007).
\bibitem{Aichhorn}M. Aichhorn, M. Hohenadler, C. Tahan, and P. B. Littlewood, Phys. Rev. Lett. \textbf{100}, 216401 (2008).
\bibitem{Cho}J. Cho, D. G. Angelakis, and S. Bose, Phys. Rev. Lett. \textbf{101}, 246809 (2008).
\bibitem{Carusotto}I. Carusotto, et al., Phys. Rev. Lett. \textbf{103}, 033601 (2009).
\bibitem{Schmidt0}S. Schmidt and G. Blatter, Phys. Rev. Lett. \textbf{103}, 086403 (2009).
\bibitem{Jens} J. Koch and K. L. Hur, Phys. Rev. A \textbf{80}, 023811 (2009).
\bibitem{Hartmann3}M. J. Hartmann, Phys. Rev. Lett. \textbf{104}, 113601 (2010).
\bibitem{Tomadin}A. Tomadin, et al., Phys. Rev. A \textbf{81}, 061801(R) (2010)
\bibitem{Tomadin2}A. Tomadin and R. Fazio, J. Opt. Soc. Am. B \textbf{27}, A130(2010).
\bibitem{Ciccarello}F. Ciccarello, Phys. Rev. A \textbf{83}, 043802 (2011).

\bibitem{Raimond} J. M. Raimond, M. Brune and S. Haroche, Rev. Mod. Phys. \textbf{73}, 565 (2001).
\bibitem{Birnbaum} K. M. Birnbaum, et al., Nature(London) \textbf{436}, 87 (2005).
\bibitem{Reithmaier} J. P. Reithmaier et al., Nature(London) \textbf{432}, 197 (2004)
\bibitem{Hennessy} K. Hennessy, et al., Nature(London) \textbf{445}, 896 (2007).


\bibitem{Irish} E. K. Irish, C. D. Ogden, and M. S. Kim, Phys. Rev. A \textbf{77}, 033801 (2008).
\bibitem{Ogden} C. D. Ogden, E. K. Irish, and M. S. Kim, Phys. Rev. A \textbf{78}, 063805 (2008).
\bibitem{zheng}Z. B. Yang, H. Z. Wu, W. J. Su, and S. B. Zheng, Phys. Rev.  A \textbf{80}, 012305 (2009).
\bibitem{Irish1} E. K. Irish, Phys. Rev. A 80, 043825 (2009).
\bibitem{zhang} K. Zhang and Z. Y. Li, Phys. Rev. A \textbf{81}, 033843 (2010).
\bibitem{Ferretti} S. Ferretti and L. C. Andreani, Phys. Rev. A \textbf{82}, 013841 (2010).
\bibitem{Schmidt} S. Schmidt, D. Gerace, A. A. Houck, G. Blatter, and H. E. T$\ddot{u}$reci, Phys. Rev. B \textbf{82}, 100507(R) (2010).
\bibitem{Guo} X. Y. Guo and Z. Z. Ren, Phys. Rev. A \textbf{83}, 013809 (2011).
\bibitem{Kanp} M. Knap, E. Arrigoni, and W. von der Linden, J. H. Cole, Phys. Rev. A \textbf{83}, 023821 (2011).

\bibitem{Scheibner} M. Scheibner et al., Nat. Phys. \textbf{3}, 106 (2007).
\bibitem{Schneble} D. Schneble et al., Science \textbf{300}, 475 (2003).

\bibitem{Nicolosi}  S. Nicolosi, et al., Phys. Rev. A \textbf{70}, 022511 (2004).
\bibitem{Wang} H. Wang, S. Q. Liu, and J. Z. He, Phys. Rev. E \textbf{79}, 041113 (2009).
\bibitem{Li} P. B. Li, Y. Gu, Q. H. Gong, and G. C. Guo, Phys. Rev. A \textbf{79}, 042339 (2009).
\end{thebibliography}
\end{document}